\documentstyle[prl,aps,preprint]{revtex}
\newcommand{\eq}{\begin{equation}}
\newcommand{\ee}{\end{equation}}
\newcommand{\eqa}{\begin{eqnarray}}
\newcommand{\eea}{\end{eqnarray}}
\def\jp{{J_\perp}}
\def\jpc{{J_\perp^c}}
\def\ajpc{{\vert J_\perp^c\vert}}

\def\ajps{{\vert J_\perp^*\vert}}
\def\s{\sigma}
\def\zp{{z_\perp}}
\def\bs{{\bf S}}
\def\br{{\bf r}}
\def\sz{S^z}
\def\sp{S^+}
\def\ag{{{\bar g}_{d+1}}}
\def\agt{{{\bar g}_{2+1}}}
\def\ajp{\vert\jp\vert}
\def\bq{{\bf q}}
\def\bqp{{\bf q_\perp}}
\def\bQ{{\bf Q}}
\def\bQp{{\bf Q_\perp}}
\def\chip{{\chi_\perp}}
\def\dbar{{\bar\Delta}}
\def\vbar{{\bar v}}
\def\bn{{\bf n}}
\def\brp{{\bf r^\prime}}
\def\lp{\Lambda_\perp}
\def\kp{{\bf k_\perp}}
\def\sm{{NL$\sigma$M}}
\begin{document}
\draft

\title{Weakly Coupled Antiferromagnetic Quantum Spin Chains}
\author{Ziqiang Wang}
\address{Department of Physics, Boston College, Chestnut Hill, MA 02167}
\maketitle

\begin{abstract}
Quasi-one-dimensional quantum antiferromagnets formed by
a $d$-dimensional hypercubic lattice of weakly coupled spin-1/2 
antiferromagnetic Heisenberg chains are studied by
combining exact results in one-dimension and renormalization group
analyses of the interchain correlations.
It is shown that $d$-dimensional magnetic long-range order develops 
at zero-temperature for infinitesimal
antiferromagnetic or ferromagnetic interchain couplings.
In the presence of weak bond alternations, the order-disorder
transition occurs at a finite interchain coupling.
Relevances to the lightly doped quantum antiferromagnets and
multi-layer quantum Hall systems are discussed.
\end{abstract}
\newpage
Low-dimensional quantum antiferromagnets (AFM) exhibit many remarkable
properties. In strictly one-dimension, transitions into ordered states
with broken symmetry is absent. For the nearest-neighbor
Heisenberg spin-$S$ chains, the low-lying excitations are gapless 
spin-1/2 quanta (spinons) for half-odd-integer $S$, whereas
a finite energy gap exists for integer $S$ \cite{hf}. 
This profound difference is captured in the effective 
O(3) nonlinear $\s$-model (\sm) description by the value of the 
topological angle ($\theta=2\pi S$) in 
the $1+1$-dimensional action \cite{hf}.

In two-dimensions, spatially isotropic
Heisenberg AFM on an unfrustrated lattice is proven
rigorously to N\'eel order in the ground state for $S\ge1$ \cite{kls}. 
While no such proof exists, it is widely believed that 
it is the case for $S=1/2$ as well. Indeed for $d\ge2$, the
long-wavelength, low-energy physics governing the interactions
between the spin-waves in the ordered phase can be described by 
a $d+1$-dimensional \sm \cite{chn}.

In this paper, we study quasi-one-dimensional quantum
AFM, and in particular, the disorder-order transition 
associated with the dimensional crossover.
Specifically, we consider $S=1/2$ 
antiferromagnetic (AF) spin chains with Heisenberg symmetry, 
arranged in a $d$-dimensional hypercubic lattice, 
and weakly-coupled by AF or ferromagnetic (FM)
interchain exchange couplings $\jp$. 
Finite intrachain bond alternations are included to study 
the competition between magnetic order and dimerization.
Our strategy is as follows.
First, interchain coupling is considered at a mean-field level
\cite{schulz,chainmf} in order to treat the important correlations that
first develop along the strongly coupled chain-direction. 
The resulting effective one-dimensional theory 
is transformed into the massive Thirring model whose
exact Bethe ansatz solution \cite{bt} is used to obtain the
static and dynamical quantities.
Then, we go beyond the mean-field theory, and show that
the order parameter fluctuations 
can be described by an anisotropic $d+1$-dimensional \sm.
Renormalization group (RG) analyses are carried out
to show that in the ground state,  $d$-dimensional magnetic long-range 
order occurs for infinitesimal interchain coupling $\vert\jp\vert>0$.
In the presence of bond alternation, long-range order develops when 
$\jp$ exceeds a finite critical value.
Aside from obvious applications to real insulating compounds behaving
as weakly coupled AF spin-1/2 chains
at low-temperatures, we will discuss
the implications of our results on the magnetic properties of
underdoped insulating cuprates and point
out the relevance to multi-layer quantum Hall structures.

The starting Hamiltonian of the system is given by
\eq
H=J\sum_{i,r}\left[1+\delta(-1)^i\right]\bs_{i,\br}\cdot\bs_{i+1,\br}
+\jp\sum_{i,\br,\mu}\bs_{i,\br}\cdot\bs_{i,\br+\mu},
\label{h}
\ee
where $\bs_{i,\br}$ is the spin-1/2 operator at lattice site
$(i,\br)$ with $i$ and $\br$ labeling the sites in the chain
($z$) and transverse to the chain ($\br_\mu$) directions,
$\mu$ is summed over the $\zp=2(d-1)$ nearest-neighbors
in the transverse directions. 
The intrachain exchange coupling is AF with
alternating strengths $J(1\pm\delta)>0$, whereas the interchain coupling can 
be either AF ($\jp>0$) or FM ($\jp<0$). We are interested in the 
case where $\delta, \vert\jp\vert/J\ll1$.

The mean-field decoupling of the interchain term in Eq.~(\ref{h})
with respect to AF order in the $z$-direction in spin space
leads to an effective Hamiltonian,
\eq
H_{1D}=J\sum_{i}\left[1+\delta(-1)^i\right]\bs_i\cdot\bs_{i+1}
-h\sum_i(-1)^i\sz_i,
\label{h1d}
\ee
plus a constant term $H_0=\zp N_{s}\ajp m_0^2/2$.
Here $N_s$ is the number of sites along the chain
and $m_0=(-1)^i\langle \sz_i\rangle$ is the staggered magnetization.
Eq.~(\ref{h1d}) describes a 1D AFM in a self-consistent 
staggered magnetic field $h=\zp\ajp m_0$. 

Next, we perform a standard
Jordan-Wigner transformation $\sz_i=\psi_i^\dagger\psi_i-1/2$, 
$\sp_i=\psi_i^\dagger\exp(i\pi\sum_{j=1}
^{i-1}\psi_j^\dagger\psi_j)$. In terms of the usual left (L) and right (R)
moving fermionic fields $\psi_L$ and $\psi_R$,
the resulting theory in the continuum limit is given by
\eqa
H_{1D}^\prime&=&\int\!dz\biggl[-iv(\psi_L^\dagger\partial_z\psi_L-
\psi_R^\dagger\partial_z\psi_R)+2g\psi_L^\dagger\psi_R^\dagger\psi_R\psi_L
\nonumber \\
&-&h(\psi_L^\dagger\psi_R+\psi_R^\dagger\psi_L)
+i\delta J(\psi_L^\dagger\psi_R-\psi_R^\dagger\psi_L)\biggr].
\label{h1dp}
\eea
It is well known that the values of $v$ and $g$
obtained in the naive continuum limit are not correct in the Heisenberg
limit. However, a comparison to the exact excitation spectrum of
the Hamiltonian in Eq.~(\ref{h1d}) at $h=\delta=0$ \cite{baxter} leads
to $v=\pi Ja/2$, where $a$ is the lattice constant set to unity hereafter.
The terms proportional to $h$ and $\delta J$ in Eq.~(\ref{h1dp}) are
easily seen in their bosonized forms to be relevant operators 
of dimension $x=1/2$ \cite{affleckles}. These competing (AF order 
v.s. dimerization) interactions will induce a mass gap ($\Delta$) 
with scaling exponent $1/(2-x)$. Thus, 
$\Delta/v\propto(h/v, \delta)^{2/3}$.

Under a global chiral rotation:
$\psi_L\to\exp(i\theta/2)\psi_L$, $\psi_R\to\exp(-i\theta/2)\psi_R$ 
with $\theta={\rm tan}^{-1}\delta J/\Delta_0$ and $\Delta_0^2=h^2
+\delta^2J^2$, terms proportional to $h$ and $\delta J$ transform
into $-\Delta_0(\psi_L^\dagger\psi_R+\psi_R^\dagger\psi_L)$.
The resulting Hamiltonian
is then identical to the massive Thirring model with bare mass $\Delta_0$
and interaction $g$, which was solved by Bethe ansatz \cite{bt}. 
Following Ref.\cite{schulz}, we obtain $g=2v=\pi J$ and
%
the ground state energy gain per site due to $m_0$,
\eq
\Delta E=\zp\ajp{m_0^2/2}-J({7/10\pi^{1/3}})
(\Delta_0/ J)^{4/3}.
\label{energy}
\ee
The self-consistent value for the staggered magnetization is obtained by
minimizing $\Delta E$ with respect to $m_0$. We find
a critical value for the interchain coupling,
\eq
\ajpc=({15\pi^{1/3}/14\zp})J\delta^{2/3},
\label{jc}
\ee
which separates an AF phase for $\ajp>\ajpc$ where the
renormalized mass gap $\Delta=(14\sqrt{3}/5\pi)\zp\ajp$ and
\eq
m_0=\left(\zp\over\pi\right)^{1/2}\left({14\over15}\right)^{3/2}\left\vert
{\jp\over J}\right\vert^{1/2}\left(1-\left\vert{\jp_c\over
  \jp}\right\vert^3\right)^{1/2}, 
\label{m0}
\ee
from a dimerized phase for $\ajp<\ajpc$ where
$\Delta_{\rm dis}=(3\sqrt{3}/\pi^{2/3})J\delta^{2/3}$ and $m_0=0$.

The mean-field theory predicts a N\'eel temperature $T_N\propto\zp\jp$ in
the ordered phase. While this can be correct when the coordination
number $\zp$ is large, it obviously contradicts the Mermin-Wagner
theorem, {\it i.e.} AF long-range order should not be possible at
any finite temperature in $d=2$. It is thus necessary to go beyond
the mean-field theory and include the order parameter fluctuations.
To this end, we turn to the dynamical spin correlations in
the ordered phase. 
Note that the ordering wavevector $\bQ=(\bQp,\pi)$ where
$\bQp=(\pi,\pi,\dots)$ for $\jp>0$ and $\bQp=(0,0,\dots)$ for $\jp<0$.
Since the translation symmetry is broken, the uniform and the staggered
components of the spins are coupled by umklapp scattering with
momentum transfer $\bq\to\bq+\bQ$. The transverse susceptibility 
in the random phase approximation is therefore
given by a $2\times2$ matrix relation,
\eq
\chi(\bq,\omega)=\chi^0(q_z,\omega)\left[{\bf1}-\ajp
  f(\bqp)\chi^0(q_z,\omega)\right]^{-1},
\label{chi}
\ee
where $\bq=(\bqp,q_z)$, and $f(\bqp)=\sum_\mu\exp(i\bqp\cdot
{\bf\mu})$. Using the equations of motion obtained for $H_{1D}$,
and the Lorentz-invariance of $H_{1D}^{\prime}$ valid at
low energies $\omega\ll J$, it is straightforward to show that the
components of the 1D susceptibility $\chi^0_{uu}$,
$\chi_{us}^0$, and $\chi_{su}^0$ are entirely determined by
$\chi_{ss}^0(q_z,\omega)$ in the long wavelength limit \cite{schulz}.
The latter has the following form,
\eq
\chi_{ss}^0(q_z,\omega)={w\over\Delta^2+v^2 q_z^2-\omega^2}
+{\cal M}(\omega^2-v^2q_z^2).
\label{chi0}
\ee
Here the pole in the first term arises from the lowest energy 
triplet excitation which corresponds to an added fermion in the 
Thirring model.
The function ${\cal M}(x)$ contains the contributions from the continuum 
involving particle-hole excitations in the Thirring model.
The latter has a threshold singularity at and a vanishing spectral weight
below $\omega=2\Delta$. Thus $\chi(\bq,\omega)$ in Eq.~(\ref{chi}) is
dominated by the collective excitations for $\omega<2\Delta$.
We will neglect the contributions from ${\cal M}(x)$ in Eq.~(\ref{chi0}),
which is equivalent to the single mode approximation (SMA). The constant $w$ in
Eq.~(\ref{chi0}) is then fixed at $w=\Delta^2/\zp\ajp$ by the condition
$\chi_{ss}^0(0,0)=1/\zp\ajp$.

Solving for $\chi(\bq,\omega)$ in Eq.~(\ref{chi}) using the SMA, we
obtain the staggered transverse susceptibility
\eq
\chi_s(\bq,\omega)={\Delta^2\over\zp\ajp}{1-h^2/\Delta^2\over
\omega_\bq^2-\omega^2},
\label{chis}
\ee
where $\omega_\bq$ is the gapless spin wave dispersion (Goldstone
modes), $\omega_\bq^2 =\dbar^2(1-f(\bqp)/\zp)+\vbar^2 q_z^2$
in terms of the weakly modified
mass gap $\dbar=\Delta\sqrt{1-h^2/\Delta^2}$ and velocity
$\vbar=v\sqrt{1-h^2/\Delta^2}$ due to the interchain correlations \cite{note}.
The uniform static susceptibility,
\eq
\chip={h^2/\Delta^2\over\zp\ajp}{1\over 1-h^2/\Delta^2}.
\label{chip}
\ee
In the limit $\ajp\to0$ and $\delta=0$,
$\chip\simeq1.07(1/\pi^2 J)$. The close agreement of the latter 
with the exact 1D result $(1/\pi^2J)$ 
suggests that the SMA is a rather
accurate description. When $\delta\ne0$, $\chip$ and thus the
spin stiffness vanishes linearly as $\ajp\to\ajpc$.

The interactions between the AF spin waves 
can be described by the O(3) quantum \sm \cite{chn}.
From the susceptibilities derived above, the $d+1$-dimensional
Euclidean action is given by (setting $\hbar=1$),
\eq
{{\cal S}_0}={1\over2g}\sum_{\langle\br,\brp\rangle}\int\!
dz\int_0^{\beta\Lambda\vbar} \! \! d\tau\biggl[
(\partial_z\bn_\br)^2+(\partial_\tau\bn_\br)^2
+{R\over\Lambda^2}
\vert\bn_\br-\bn_\brp\vert^2\biggr].
\label{s0}
\ee
Here $\bn_\br(z,\tau)$ is a three-component unit-modulus
vector field. It represents the local orientation of the AF order parameter.
The discrete sum runs over the neighboring lattice sites in the
transverse directions. As usual, $\beta=1/k_B T$, and $\Lambda$
is a spatial cutoff 
at which the coupling constant $g=\vbar/\rho_s^0$ and $\rho_s^0=\chip\vbar^2$.
The anisotropy is contained in $R=\Delta^2/\zp v^2\simeq0.97\zp(\ajp/J)^2
\ll1$. In terms of the unit vector field, the transverse
spin susceptibility $\chi_s(\bq,\omega)=m_0^2\langle n^+(\bq,\omega)
n^-(\bq,\omega)\rangle$ and the staggered magnetization
$m=m_0\langle n^z \rangle \vert_{T=0}$.

The RG analysis of ${\cal S}_0$ is subtle. Let us consider the
case when $\delta=0$, {\it i.e.} for vanishing bond alternation.
Notice that we did not keep track of the topological term explicitly
since at $\theta=\pi$ it does not renormalize under the RG.
The effect of the latter is however crucial
for the renormalization of the coupling constant $g$ in the
$1+1$-dimensional sector in the limit $R\to0$ \cite{affhal}.
In the presence of the topological term,
$g$ flows to a finite fixed point value $g(\infty)$
and the correlation length is infinite,
whereas $g\to\infty$ and the system
develops a finite correlation length $\xi_\s/\Lambda\approx
e^{2\pi/g}$ in its absence.
Thus, during the RG transformation of Eq.~(\ref{s0}), 
if the anisotropy is large enough
such that $\sqrt{R}/\Lambda\ll1/\xi_\s$, further renormalization
using the $d+1$-dimensional RG cannot eliminate this finite correlation
length, the correct treatment of the $S=1/2$ system must include
the effect of the topological term. On the other hand, in the opposite
limit where $1\gg\sqrt{R}/\Lambda\gg1/\xi_\s$, the long-wavelength
physics is essentially controlled by the $d+1$-dimensional RG
and the topological term would not make a qualitative difference
in the ordered phase.
Below, we consider the two situations separately.

For $\sqrt{R}\ll e^{-2\pi/ g}$, we follow the analysis of
dimensional crossover \cite{affhal,stanley}.
Since $R$ is exponentially small, the RG in the $1+1$-dimensional
sector can be performed independently by integrating out high-momentum 
modes until the effective couplings become comparable
in all directions at a larger cutoff $\lp$, {\it i.e.}
when $R/\Lambda^2g\approx1/\lp^2g(\lp)$. The
$d+1$-dimensional RG is switched on thereafter.
For large $\lp$, the coupling $g(\lp)$ flows
towards its limiting fixed point value which is of order one.
Thus, the crossover length scale is $\lp\approx \Lambda/\sqrt{R}$.
Since the scaling dimension
of the $\bn$-field is zero in the ordered phase, there is no need
to rescale the latter in the interchain term in Eq.~(\ref{s0}).
Taking the continuum limit in the transverse directions
by absorbing the cutoff $\lp^{-2}$ into defining the
derivatives, Eq.~(\ref{s0}) is reduced to a continuous, isotropic
action at an isotropic cutoff $\lp$,
\eq
{{\cal S}_1}={1\over2g_{d+1}}\int\!d^{d-1}\br\int\!
dz\int_0^{\beta\lp\vbar} \! d\tau\bigl[(\partial_\br\bn)^2
+(\partial_z\bn)^2+(\partial_\tau
\bn)^2\bigr],
\label{s1}
\ee
where $g_{\rm d+1}=g(\lp)\lp^{d-1}$ is the bare coupling constant
for the $d+1$-dimensional RG. For $R\to0$, $g(\lp)\to g(\infty)$.
Thus, the stability of the AF ordered
state at infinitesimal $R$ is determined by whether $g_{d+1}$ 
is smaller than the critical coupling $g_{d+1}^c$ of the 
$d+1$-dimensional RG. The latter predicts a $T=0$ fixed
point at $g_{d+1}^c/\lp^{d-1}=(d-1)2^d\pi^{d/2}\Gamma(d/2)$ to
one-loop order \cite{chn}.

Since the exact solution of $H_{1D}$ in Eq.~(\ref{h1d}) 
describes the fixed point physics in the $1+1$-dimensional sector,
$g(\infty)=1/\chip\vbar$ can be calculated
using the results in Eqs.~(\ref{chis}) and (\ref{chip}).
(i) For $\delta=0$, 
$g(\infty)=b[1-(2/b\pi)\zp\ajp/J]^{1/2}$ with $b=810/14\pi^2$, which
approaches the value $g(\infty)=5.86$ from below as $\ajp\to0$.
Thus, the condition for AF order, $\ag \equiv g_{d+1}(\lp)/g_{d+1}^c<1$, 
is satisfied and improved further with increasing $\zp\ajp$.
This one-loop result is consistent with the numerical series expansion
analysis in $d=2$ for AF coupled chains \cite{affhal}.
We therefore conclude that in the absence of bond alternation,
long-range order develops for infinitesimal AF or FM interchain 
couplings in $d\ge2$. The physical origin of this behavior should be
traced back to the gapless power-law correlations in 
the spin-1/2 Heisenberg chain.
For $d=2$, a finite temperature fixed point does not exist in the \sm,
the ordered phase is stable only at $T=0$. This corrects the naive
mean-field prediction of a finite $T_N$.
(ii) When $\delta$ is finite, $\ajpc\ne0$ in 
Eq.~(\ref{jc}). For $\zp\ajp/J\ll1$, we find $g(\infty)\simeq
b(1-\vert\jpc/\jp\vert^3)^{-1}$ which diverges as
$\ajp$ is reduced toward $\ajpc$, indicating a transition into
a disordered phase with finite dimerization. 
The critical coupling is at a $\ajps$ ($> \ajpc$ predicted in the
mean-field theory) where $\ag=1$.
To one-loop order, $\ajps=\ajpc(1-\ag^{\delta=0})^{-1/3}$. 
The renormalized spin stiffness
vanishes on approaching the transition according to 
$\rho_s=\rho_s^0(1-\ag)$.
Since $\langle (n^z)^2\rangle=(1-\ag)$, the renormalized
staggered magnetization $m^2=m_0^2(1-\ag)$.

In the above discussion, the coupling constant $g$
defined at cutoff $\Lambda$ in Eq.~(\ref{s0}) was taken to be
close to the (order one) fixed point value of $H_{1D}$.
From the point of view of the effective {\sm},
this does not have to be the case. If $g\ll1$, 
the independent $1+1$-dimensional RG 
should be replaced by the $d+1$-dimensional RG once the reduced anisotropy
is in the range $1\gg\sqrt{R^\prime}\gg e^{-2\pi/g(\Lambda^\prime)}$
at a cutoff $\Lambda^\prime=\Lambda\sqrt{R^\prime g/R
g(\Lambda^\prime)}$. 
Further analysis then belongs to the second parameter regime
which is also relevant for a large-S AFM where $g\sim2/S$.

For $1\gg\sqrt{R}\gg e^{-2\pi/g}$,
because $R$ is no longer exponentially small,
the $1+1$-dimensional RG does not transform independently.
Instead, one should treat the $d+1$-dimensional RG
with anisotropic 
cutoffs and expect 
the residual anisotropy to persist down to the $d+1$-dimensional fixed
point. Taking the continuum limit in the transverse directions 
at cutoff $\lp=\Lambda/\sqrt{R}$ in
Eq.~(\ref{s0}) leads to an action like the one in Eq.~(\ref{s1})
with bare coupling $g_{d+1}=g\lp^{d-1}$.
The important difference is that the momentum cutoffs are now
anisotropic, $\vert\kp\vert< k_\perp^m=\pi/\lp$, $\vert k_z\vert
<k_z^m=\pi/\Lambda$. We have carried out the
momentum shell RG by integrating out modes in the
high-momentum layers of the $d$-dimensional box in $k$-space with
$k_\perp^m > \vert \kp\vert >k_\perp^m e^{-l}$ 
and $k_z^m>\vert k_z\vert >k_z^m e^{-l}$, where
$e^l$ is the length rescaling factor. At $T=0$, the
one-loop RG equation is $dg_{d+1}/dl=(1-d)g_{d+1}
+g_{d+1}^2/g_{d+1}^c(R)$. The critical coupling $g_{d+1}^c(R)$ depends
on the anisotropy $R$. For $d=2$, we find,
\eq
g_{2+1}^c(R)=2\pi\lp\sqrt{R}\left[\ln(\sqrt{1+R}+\sqrt{R})
+\sqrt{R}\ln(\sqrt{1+1/R}+1/\sqrt{R})\right]^{-1}.
\label{gc}
\ee
For $R\ll1$, $g_{2+1}^c(R)\approx2\pi\lp/\ln(2/\sqrt{R})$.
The ratio of the bare coupling to the fixed point value
is thus $\agt(R)\approx (g/2\pi)\ln(2/\sqrt{R})\ll1$ in
this parameter regime. In fact, for all $d\ge2$
one can show $g_{d+1}^c(R)\propto\pi\lp^{d-1}/\ln(2/\sqrt{R})$
such that $\ag(R)\ll1$. Thus the conclusion is again that
the weakly-coupled spin chains are in the ordered phase.
When $\delta\ne0$, the uniform susceptibility $\chip$ decreases with
decreasing $\ajp$.
Thus $g$ increases until ${\sqrt R}\sim e^{-2\pi/g}$.
The ordered phase becomes unstable at a critical value $R_c$ defined by
$\ag(R_c)=1$ where a transition into the dimerized phase takes place. 
Close to the transition, $\rho_s=\rho_s^0\sqrt{R}(1-\ag(R))$ 
and $m^2=m_0^2(1-\ag(R))$.

The finite temperature properties close to the transition are described
by the scaling behavior of the $d+1$-dimensional
quantum \sm. The correlation length ($\xi$) at low-temperatures
can be obtained by integrating the one-loop RG equations
\cite{chn}.
For $d=2$,  
in the quantum critical regime, 
$\xi=(\hbar\vbar/k_BT)2\pi\Lambda/g_{2+1}(R_c)$ at $\agt(R_c)=1$,
where $\vbar$ is the effective spin-wave
velocity along the chain direction.
For $R_c\ll1$, 
$\xi\propto(\hbar\vbar/k_BT)\sqrt{R_c}\vert\ln\sqrt{R_c}\vert$. 
In the renormalized classical regime,
$\xi\propto\sqrt{R}\vert\ln\sqrt{R}\vert(\hbar\vbar/k_BT)
\exp{(2\pi\rho_s/k_B T)}.$
Thus the effects of anisotropy enter the prefactors 
of these universal functions.


Recently, it was proposed that the destruction of AF long-range
order in the lightly-doped insulating cuprates may be explained by 
the dimensional crossover in an effective low-energy theory of a
2D Heisenberg AFM with increasing anisotropy \cite{ch}.
The important question is whether the effective spin quantum number
of the AFM corresponds to an integer or a half-odd-integer.
Since the effective theory is motivated by the physics of the 
striped phases \cite{theory,exp,borsa}
with a large correlation length compared the
distance between the stripes, the effective spin presumably depends
on the number of spin-1/2 chains between the
neighboring stripes. The latter
is, unfortunately, not known precisely in the lightly-doped regime.
It was shown in Ref.~\cite{ch},
that for {\it integer} effective spins (appropriate for
the case of an even number of spins between the stripes),
AF long-range order disappears below a {\it finite} 
critical interchain coupling related to Eq.~(\ref{gc}).
In the case where the effective
spin corresponds to a half-integer, our results show that AF long-range order
is stable for {\it arbitrarily small} interchain couplings in
agreement with previous theoretical work \cite{affhal}.
In order to destroy AF order in this case,
small dimerization is necessary in addition to anisotropy. 
Note that 
the amount of dimerization (bond alternation) necessary
to disorder the system and open up a spin gap is $\delta_c\propto
(\jp/J)^{3/2}$ which can be very small, whereas
$\delta_c\approx0.3$ on a dimerized square lattice Heisenberg AFM
without spatial anisotropy \cite{imada}.

The properties of the $S=1/2$ AF spin chains can provide useful
insights into the integer quantum Hall transitions.
In a single layer quantum Hall structure (QHS), the latter is in 
the universality class of the dimerization (spin-Peierls) 
transition of an SU(2n) AF quantum spin chain in the limit $n\to0$
\cite{lw}. The double-layer (or spin Landau level mixing) case 
corresponds to two FM coupled spin chains \cite{wlw}.
These results suggest qualitatively similar phase structures
in the $n=1$ and $n\to0$ cases.
The difference in the universal properties of the phase 
transitions can be summarized by the changes in the critical exponents
with $n$. A large number of coupled multi-layer QHS
corresponds naturally to an array of FM coupled spin chains, 
where the interchain coupling originates from tunneling between the
layers.
The quasi-1D spin chains may order for infinitesimal $\ajp>0$,
as we have shown for $n=1$. The resulting spin wave spectrum just
describes the diffusive modes that would appear
in a disordered metal, suggesting the formation of
a metallic phase 
between the insulator/quantum Hall states, consistent
with recent numerical simulations \cite{chalker}.
The corresponding phase transitions are thus
in the universality class of the $U(2n)/U(n)\times U(n)\vert_{n\to0}$
\sm. Interestingly, the latter also describes the 3D Anderson transition
in the presence of time-reversal symmetry breaking.

The author thanks A.~H. Castro Neto for a discussion.
This work was supported in part by an award from Research Corporation.

\vspace*{\fill}{

}

\begin{references}
\bibitem{note} In deriving these results, we have dropped 
negligible contributions of ${\cal O}(\delta^2\ajp/J)$
to the spin wave dispersion.
\end{references}
\end{document}